\documentclass{moriond}

\usepackage{cite}
\usepackage[normalem]{ulem}
\def\be{\begin{equation}}
\def\ee{\end{equation}}
\def\bea{\begin{eqnarray}}
\def\eea{\end{eqnarray}}

\newcommand{\Photo}{}

\begin{document}
\vspace*{4cm}
\title{Gravitational waves and the Higgs mass}

\author{ Rui Santos}

\address{Centro de F\'{\i}sica Te\'{o}rica e Computacional,
   Faculdade de Ci\^{e}ncias,\\
   Universidade de Lisboa, Campo Grande, Edif\'{\i}cio C8
  1749-016 Lisboa, Portugal.}

\address{ISEL - Instituto Superior de Engenharia de Lisboa,\\
  Instituto Polit\'ecnico de Lisboa
 1959-007 Lisboa, Portugal.}

\maketitle\abstracts{
One of the simplest models that allows for an electroweak (EW) strong first order phase transition (SFOPT) while
having a Higgs with the observed mass of 125 GeV is the singlet extension of the Standard Model (SM). 
We discuss two different phases of a complex singlet extension of the SM. If the singlet has a zero vacuum expectation value (VEV) 
the situation is exactly the same as in the SM with no SFOPT. However, when the singlet acquires a VEV, a SFOPT is possible,
giving rise to detectable gravitational waves (GWs) by LISA. The model also provides a dark
matter candidate. We have studied the impact of the SM particle masses on the strength of the GWs spectrum. It turns out that the variation
of the Higgs boson and top quark masses within one standard deviation of the
experimentally measured value has a noticeable impact on GWs detectability prospects.  
}

\section{Introduction}

If today's Higgs potential results from an EW SFOPT in the Higgs vacuum that occurred in the early Universe one could expect to find its remnants as primordial GWs  
that could be detected in the near future~\cite{Witten:1984rs,Hogan:1986qda,Kamionkowski:1993fg}. A SFOPT is needed to generate the observed baryon asymmetry~\cite{Sakharov:1967dj}
but the SM fails to provide it due to the large Higgs boson mass. We have studied~\cite{Freitas:2021yng} an extension of the SM by an extra gauge complex singlet that can trigger a SFOPT
provided the singlet VEV is non-zero (see also~\cite{Espinosa:2007qk, Ashoorioon:2009nf, Kakizaki:2015wua}). There is a connection between the parameters of the scalar potential, the duration and latent heat of the SFOPT 
and the peak amplitude and frequency of the associated primordial GW spectrum. Once points with a large enough peak amplitude to be detected in the forthcoming experiments were found,
two main problems were investigated: the dependence of the peak amplitude and frequency with the SM particle masses and the dependence with the singlet VEV.

\section{Models and GWs detection}

The Higgs potential is the SM potential plus a complex gauge singlet $\sigma$ with a global softly-broken $U(1)$ symmetry,
\begin{eqnarray}
\mathcal{V}_0(\Phi,\sigma) & = & \mu_\Phi^2 \Phi^\dag \Phi + \lambda_\Phi (\Phi^\dag \Phi)^2 
+ \mu_\sigma^2 \sigma^\dag \sigma + \lambda_\sigma (\sigma^\dag \sigma)^2 
+ \lambda_{\Phi \sigma} \Phi^\dag \Phi \sigma^\dag \sigma + 
\Big(\frac12 \mu_b^2 \sigma^2 + {\rm h.c.}\Big) \, ,
\label{V0-tot}
\end{eqnarray}
where $\Phi$ is the SM Higgs doublet and $ \sigma = \frac{1}{\sqrt{2}}(\phi_\sigma + \sigma_R + i \sigma_I )$ with real $\sigma_R$ and $\sigma_I$. $h$ is the Higgs boson field, which is a 
quantum fluctuation about the classical mean-field $\phi_h$ for which $\phi_{h}(T=0)\equiv v_{h} = 246$ GeV. The two scenarios to be discussed are: 
\begin{itemize}
\item \textbf{Scenario 1} -- only the doublet acquires a VEV whereas the singlet VEVs are both zero at $T=0$ \textit{i.e.}~$\phi_{\sigma}(T=0)=0$. At finite temperatures, the real component may fluctuate around a non-zero $\phi_\sigma(T)$;
\item \textbf{Scenario 2} -- both the doublet and the real part of the singlet have non-zero VEVs at $T=0$, that is, $\phi_{h,\sigma}(T=0)\equiv v_{h,\sigma}$. One of the CP-even scalar states is the SM-like Higgs ($125$ GeV). The soft breaking term explicitly breaks $U(1)_L \to Z_2$ providing a pseudo-Goldstone mass to $\sigma_I$.
\end{itemize}

The physical quantities needed to describe the GW signals originating from EW FOPTs in the early Universe are obtained from the effective potential~\cite{Quiros:1999jp,Curtin:2016urg},
\begin{equation}
V_{\rm eff}(T) = V_0 + V^{(1)}_{\rm CW} + \Delta V(T) + V_{\rm ct}\,,
\label{eff-pot}
\end{equation}
written in terms of the tree-level potencial $V_0$, the Coleman-Weinberg (CW) potential, $V^{(1)}_{\rm CW}$~\cite{Coleman:1973jx},
the one-loop thermal corrections, $\Delta V(T)$~\cite{Quiros:1999jp} and the counterterm potential $V_{\rm ct}$ (see~\cite{Freitas:2021yng}  for the complete expressions for each model).
There are three temperatures relevant to the phase transition:  the critical temperature at which the effective potential has two degenerate minima;
the nucleation temperature, $T_n$, for which   the true vacuum emerges and the FOPT becomes efficient, which means that  the transition probability is of order one per unit Hubble time and Hubble volume;
the percolation temperature,  $T_*$, at which at least $34\%$ of the false vacuum has tunnelled into the true vacuum~\cite{Ellisupdated}. 
The relevant parameters for the phase transition are:
\begin{itemize}
\item The strength of the phase transition, $\alpha$, related to the latent heat released in the FOPT at the bubble percolation temperature $T_*$~\cite{Hindmarsh:2015qta,Hindmarsh:2017gnf}
\begin{equation}
\alpha = \frac{1}{\rho_\gamma} 
\left[ V_i - V_f - \frac{T_*}{4} \left( \frac{\partial V_i}{\partial T} - 
\frac{\partial V_f}{\partial T} \right) \right] \,, \qquad \rho_\gamma = g_* \frac{\pi^2}{30} T_*^4
\label{alpha}
\end{equation}
where $\rho_\gamma$ is the energy density of the radiation medium at the bubble percolation epoch written as a function of the effective number of relativistic degrees of freedom~\cite{Grojean:2006bp,Leitao:2015fmj,Caprini:2015zlo,Caprini:2019egz}. 
 $V_i$ and $V_f$ are the values of the potential before and after the phase transition.
\item
The inverse time-scale of the phase transition in units of the Hubble parameter $H$ is
\begin{equation}
\frac{\beta}{H} = T_*  \left. \frac{\partial}{\partial T} \left( \frac{\hat{S}_3}{T}\right) \right|_{T_*}\,,
\label{betaH}
\end{equation}
where $\hat{S}_3$ is the Euclidean action.
\item
We only consider the case of non-runaway nucleated bubbles following Ref.~\cite{Caprini:2019egz} to estimate the spectrum of primordial GWs. The intensity of the GW radiation grows with the ratio 
$\Delta v_\phi/T_*$,  the order parameter, where
\begin{equation}
    \Delta v_\phi = |v_\phi^f - v_\phi^i|\,, \qquad \phi = h,\sigma
\end{equation}
is the difference between the VEVs of the initial (metastable) and final (stable) phases at the percolation temperature $T_*$.
\end{itemize}

\section{Results and discussion}

\begin{figure}[h!]
\begin{center}
\includegraphics[width=0.39\linewidth]{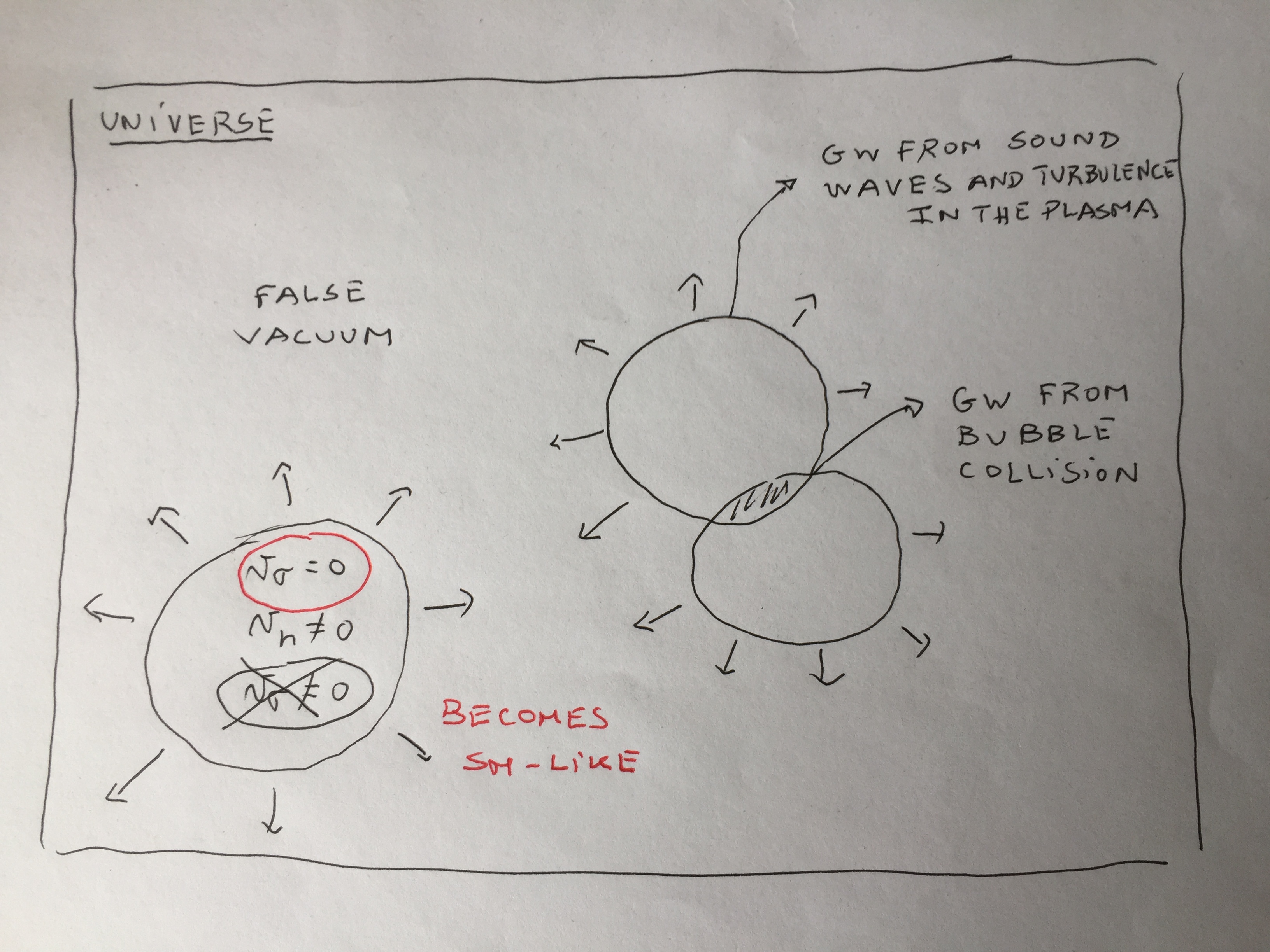}
\includegraphics[width=0.39\linewidth]{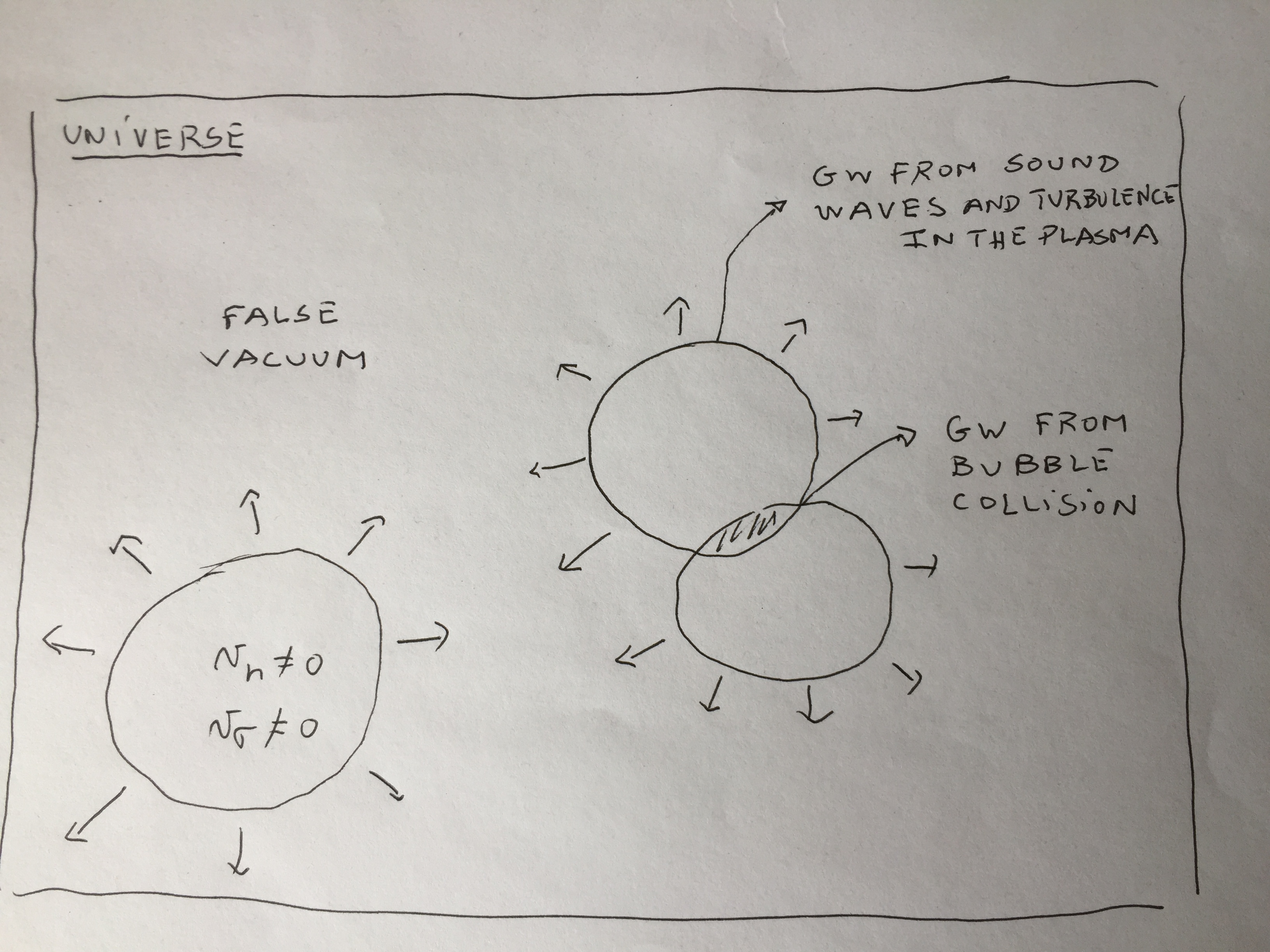}
\end{center}
\caption[]{Two pictures of the early universe. Left - the singlet has no VEV at zero temperature and no SPOPT occurs. Right - the singlet has a VEV at zero temperature and a SFOPT may occur giving rise to GWs originated from shock sound waves.}
\label{fig:radish}
\end{figure}
\begin{figure}[h!]
\centerline{\includegraphics[width=0.7\linewidth]{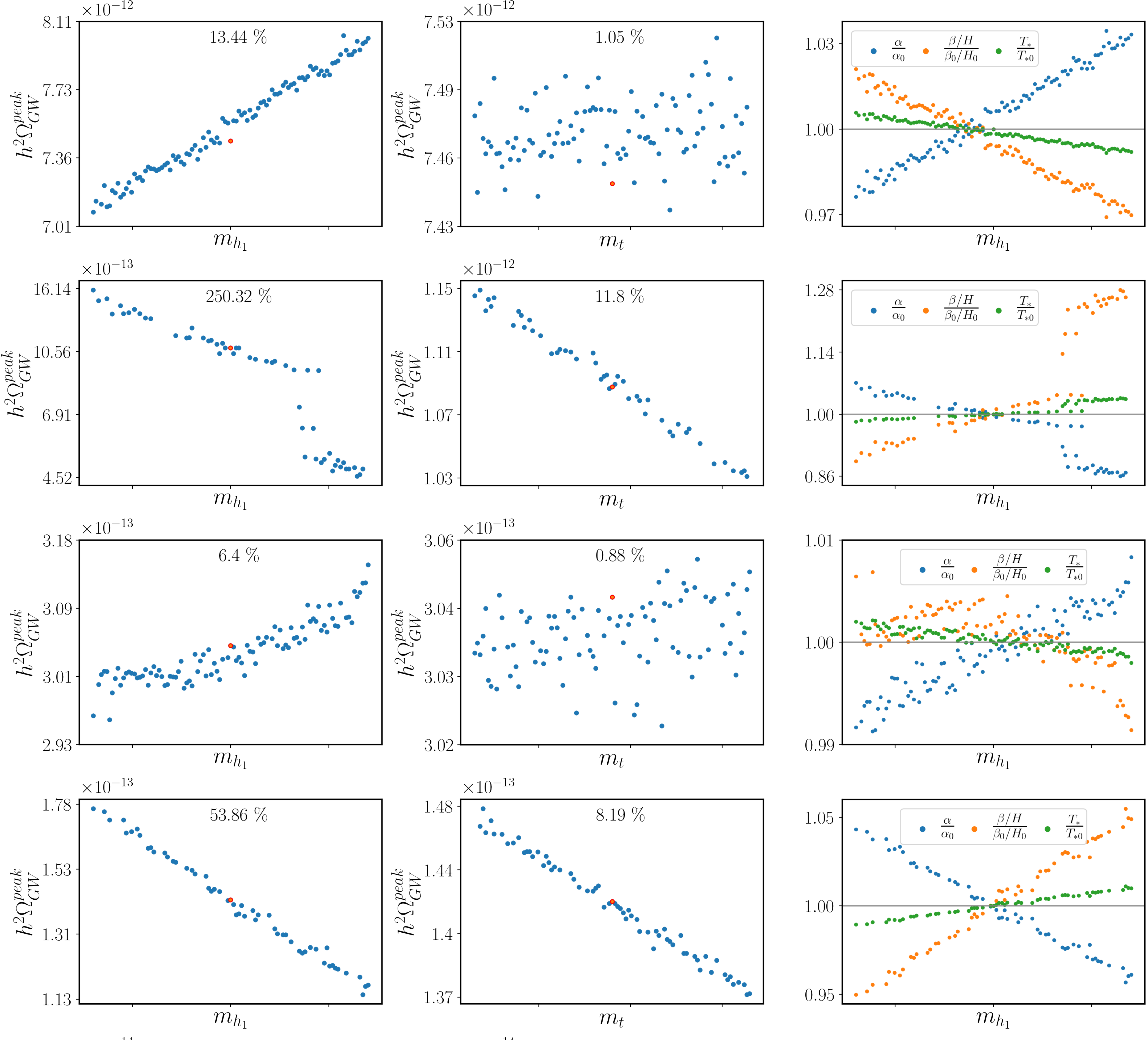}}
\caption[]{The dependence of $h^2 \Omega^{\rm peak}_{\rm GW}$ on the Higgs boson mass $m_{h_1}$ (left) and on the top-quark mass (center). In the right plot we present the corresponding variation for the parameters $\alpha$,  $\beta/H$ and $T_*$, where the subscript $0$ denotes the central values.
For each point all other parameters of the model are fixed.  The chosen original FOPT (before parameters' variation) is denoted by a red circumference. All masses are given in GeV.}
\label{FigSM}
\end{figure}
We consider only GWs originating from sound shock waves generated by the bubble's violent expansion in the early Universe. In Fig.~\ref{fig:radish}
we show two pictures of the early universe. In the left panel the singlet has no VEV at zero temperature and no SPOPT occurs, as is the case for the SM. In the right panel we show a universe 
where the singlet has a VEV at zero temperature and a SFOPT may occur giving rise to GWs.
The impact of the SM parameters in the GWs was done by first fixing the SM particle masses at their central values according to the PDG~\cite{Zyla:2020zbs} followed
by looking for points within the reach of the LISA experiment. For each point found we then varied within $1\sigma$ and one at a time each of the fermion masses, from the electron
to the top quark, the $W$ and $Z$ bosons' masses and the Higgs mass. We concluded that the only SM parameters with a meaningful impact on the 
 GW peak amplitude and frequency were the top quark mass and even more so the Higgs boson mass.
In Fig.~\ref{FigSM} we show four points within LISA's reach for the scenario with a non-zero VEV, identified by a red circle. In the left column we present the variation of the
peak amplitude with the Higgs mass in the interval from 124.96 GeV to 125.24 GeV ($1\sigma$ uncertainty), while keeping the remaining 
parameters of the SM and the ones from the dark sector constant. The maximum variation found for the peak amplitude was 250 \% for the Higgs mass. For the top quark 
mass a maximal variation of 50 \% was obtained for $m_t$ varied between 172.46 GeV and 173.06 GeV.

\section{Conclusions}

We have discussed the observation of primordial GWs originating from a SFOPT. The variation of the SM particle masses does indeed lead to noticeable differences both in the peak of the GW power spectrum
and on its frequency. The effect is more pronounced for the Higgs mass but also true for the top quark mass. The remaining masses have a very mild, if any,
impact on GW spectrum. We should underline the point that the variation of the Higgs mass within the measured experimental
error can lead to at least 250 \% magnitude change in the GW peak.

We also concluded that the exact same model but with two different phases at zero temperature leads to very different results: no VEV in the singlet means a SM-like behaviour while if the singlet has a VEV
there are good chances of detecting GWs by LISA. 

\section*{Acknowledgments}

Work supported by the Portuguese Foundation for Science and Technology (FCT) under Contracts no. UIDB/00618/2020, UIDP/00618/2020, PTDC/FIS-PAR/31000/2017 and CERN/FIS-PAR/0014/2019.

\section*{References}
\bibliographystyle{h-physrev}
\bibliography{GW.bib}

%\begin{thebibliography}{99}
%\bibitem{ja}C Jarlskog in {\em CP Violation}, ed. C Jarlskog
%(World Scientific, Singapore, 1988).
%
%\bibitem{ma}L. Maiani, \Journal{\PLB}{62}{183}{1976}.
%
%\bibitem{bu}J.D. Bjorken and I. Dunietz, \Journal{\PRD}{36}{2109}{1987}.
%
%\bibitem{bd}C.D. Buchanan {\it et al}, \Journal{\PRD}{45}{4088}{1992}.
%
%\end{thebibliography}

\end{document}